\documentclass{ethpaper}
\pdfoutput=1
\usepackage{ifpdf}
\usepackage{graphicx}
\usepackage{amssymb}
\usepackage{subfigure}
\usepackage{epstopdf}
\begin{document}
\begin{titlepage}
\ethnote{}
\title{Results on damage induced by high-energy protons in LYSO calorimeter crystals}
\begin{Authlist}
G. Dissertori, D.~Luckey, F.~Nessi-Tedaldi, F.~Pauss, M.~Quittnat, R.~Wallny
\Instfoot{eth}{Institute for Particle Physics, ETH Zurich, 8093 Zurich, Switzerland}
M.~Glaser
\Instfoot{cern}{CERN - PH Department, 1211 Geneva 23, Switzerland}
\end{Authlist}
\maketitle
\begin{abstract}
Lutetium-Yttrium Orthosilicate doped with Cerium (LYSO:Ce), as a bright scintillating crystal, is a candidate for calorimetry applications in strong ionizing-radiation fields and large high-energy hadron fluences as are expected  at the CERN Large Hadron Collider after the planned High-Luminosity upgrade. There, proton-proton collisions will produce fast hadron fluences up to $\sim5\times10^{14}/\mathrm{cm}^2$
in the large-rapidity regions of the calorimeters.

The performance of LYSO:Ce has been investigated, after exposure to different fluences of 24 GeV/c protons.
Measured changes in optical transmission as a function of proton fluence are presented, and
the evolution over time due to spontaneous recovery at room temperature is studied.

The activation of materials will also be an issue in the described environment. Studies of the ambient dose induced by LYSO and its evolution with time, in comparison with other scintillating crystals, have also been performed through measurements and FLUKA simulations.
\end{abstract}

\vspace{7cm}
\conference{submitted to Elsevier for publication in Nucl. Instr. and Meth. in Phys. Research A}

\end{titlepage}

\section{Introduction}
\label{s-INT}
The planned upgrade to High-Luminosity running of the Large Hadron Collider at CERN  (HL-LHC) will impose 
performance requirements on detectors, that are more stringent than those adopted for LHC construction two decades ago.
According to the present schedule, such an upgrade should start in 2022. 
Some of the detectors envisage an upgrade, primarily of their end caps, and materials suitable for that challenging environment need to be swiftly identified.
If adopted for electromagnetic calorimetry, inorganic scintillators will have to perform adequately in an environment where ionizing radiation levels can be as high as 65 Gy/h, and energetic hadron fluences are expected to reach integrated values of $5\times 10^{14}\;\mathrm{cm}^{-2}$.

The present work extents our earlier studies of Lead Tungstate~\cite{r-LTNIM,r-LYNIM,r-pionNIM,r-FTNIM}  and Cerium Fluoride~\cite{r-CEF3}, to explore potentially suitable scintillators.
Therein, we have shown how Lead Tungstate (PbWO$_4$) exposed to hadronic showers from high-energy
protons~\cite{r-LTNIM} and pions~\cite{r-pionNIM} experiences a
cumulative loss of Light Transmission which is permanent at room temperature,
while  no hadron-specific
change in scintillation emission~\cite{r-LYNIM} was observed. The amplitude of the damage can reach values
unsuitable for running at the HL-LHC~\cite{r-RADQ,r-IEEE12}. A microscopic investigation~\cite{r-FTNIM} has
allowed us to ascribe the dominant damage mechanism to heavy fragments arising from Lead and Tungsten fission, which
deposit a lot of energy along their short track, leaving regions within the crystal where the lattice
structure is modified and left disturbed, strained, disordered, or re-oriented.
These damage regions have different optical and mechanical properties compared to the surrounding 
crystal lattice, and  they can act as scatterers for light propagating in the crystal, thus affecting the light
transmission.

The qualitative understanding we gained of hadron damage in Lead
Tung\-state led to the prediction~\cite{r-CAL08} that such hadron-specific
damage contributions are absent in crystals  consisting only of
elements with $Z < 71$, which is the experimentally observed threshold
for fission~\cite{r-THR}, while they should be expected in crystals
containing elements with  $Z > 71$. We confirmed the first prediction with measurements~\cite{r-CEF3}
that show how hadrons in Cerium Fluoride cause a damage that recovers at room temperature, 
with none of the features present, which we observed for Lead Tungstate, thus making it an excellent candidate for HL-LHC applications.
The second prediction is confirmed by existing proton-damage measurements
in BGO~\cite{r-KOBA,r-DPF}, in Lead Fluoride and in BSO 
~\cite{r-KOR}, which all contain elements with $Z > 71$.

All this evidence leads to the guideline that for resistance to damage from energetic hadrons, materials need
to consist of light elements. In the light of all this, Cerium-doped Lutetium-Yttrium Orthosilicate (chemical formula Lu$_{2(1-x)}$Y$_{2x}$SiO$_5$:Ce or briefly LYSO), which is a commercially available scintillator, is of crucial interest, since it contains Lutetium, that with Z=71 sits right at the threshold for fission. For this reason, and for all the further characteristics described in section~\ref{s-LYSO}, this study addresses the effect of energetic hadrons on the performance of LYSO scintillating crystals. Its outcome is a highly relevant input to decisions on calorimeter technology for HL-LHC running and for further applications where an exposure to intense fluxes of energetic hadrons has to be taken into account.

\section{LYSO}
\label{s-LYSO}
Cerium-doped silicate-based crystals were recently developed for medical applications.
Initially, LSO (Lu$_{2}$SiO$_{5}$:Ce) was first investigated as a phosphor for cathode ray tube displays~\cite{r-GOM}, then rediscovered as a promising scintillator and first grown in 1989 ~\cite{r-MEL}. A mass-production was later established for LSO~\cite{r-MAS} and for LYSO ~\cite{r-COK} in the early years 2000.
LYSO is a mixed crystal, a dense, bright scintillator, which is nowadays commercially available, being industrially produced by several companies for high-precision Positron-Emission Tomography.
Its density ($\rho=7.4\; {\mathrm{g/cm}}^3$), radiation length ($X_0 =
1.14$ cm), Moli\`ere radius ($R_M = 2.07$ cm), nuclear interaction length
($\lambda_I = 20.9$ cm) and refractive index ($n = 1.82$) make it a
competitive medium for compact calorimeters. Its emission is centred
at 430 nm~\cite{r-MAO2}, with a decay time constant of 40 ns; it is rather insensitive
to temperature changes (${\mathrm{dLY/dT}}\; (20^o$C$)=-0.2\%/^o$C) as well as bright (85\% of NaI($T\ell$)) and thus suitable for high-rate, high-precision
calorimetry.

\begin{figure}[h]
\begin{center}
	{\mbox{\includegraphics[width=10cm,trim=10mm 10mm 10mm 14mm , clip]{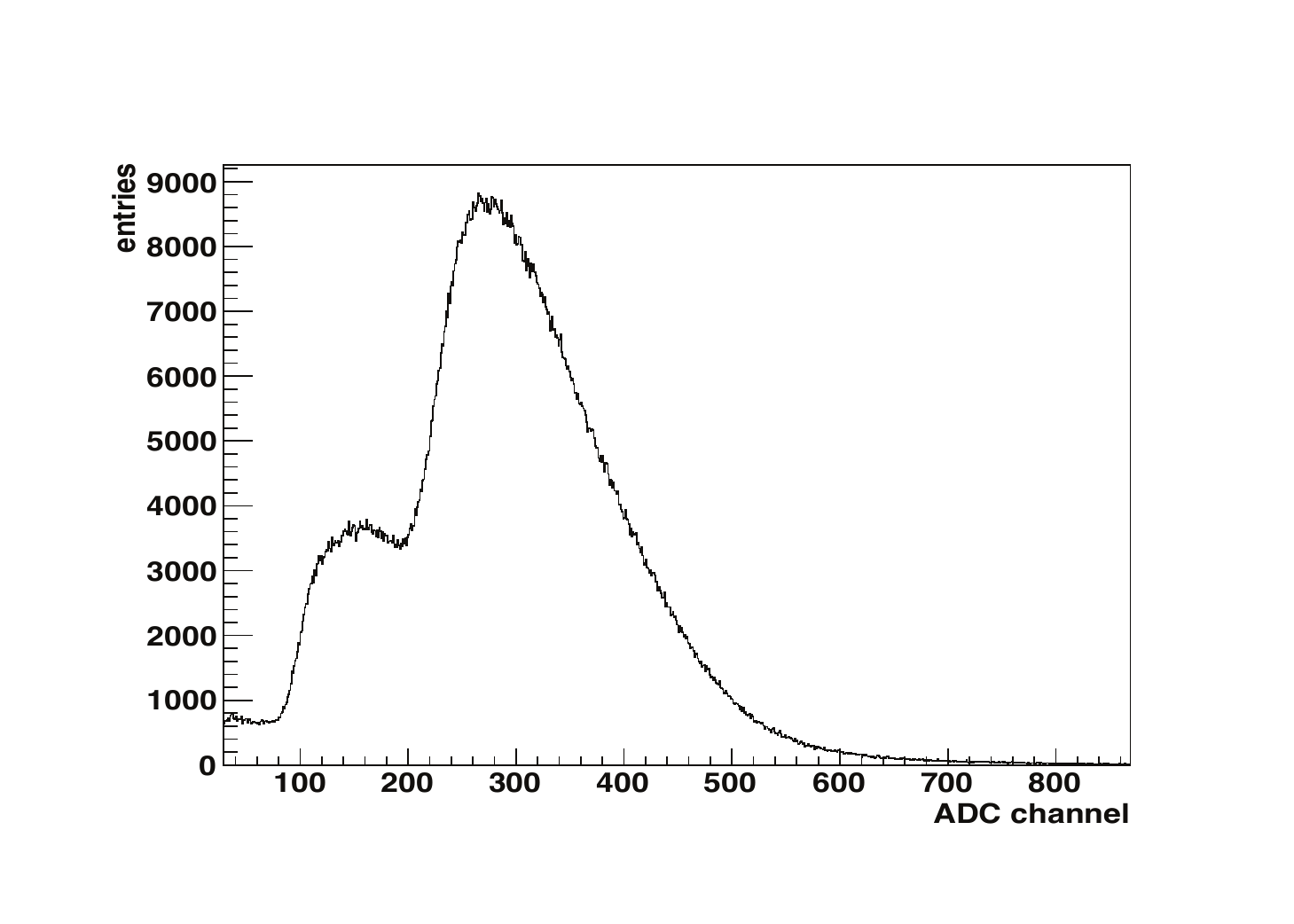}}}
\end{center}
\caption{Energy deposition spectrum from a $25\times 25\times 100\;\mbox{mm}^3$ LYSO crystal, due to decays of $^{176}$Lu. The continuous $\beta$ spectrum is distorted by discrete shifts in energy caused by the simultaneous detection of one, two or three of the photons emitted due to the de-excitation of the daughter isotope, $^{176}$Hf\label{f-SPEC}.}
\end{figure}

LYSO contains a fraction of the naturally occurring isotope $^{176}\mathrm{Lu}$, that undergoes $\beta$-decay followed by the emission of 88, 202 and 307 keV $\gamma$-rays, that cause signals which can be assimilated to phosphorescence. These can be exploited for calibration purposes, in that the energy deposition of the concomitant $\gamma$ rays, that add to the continuum due to the $\beta$-decay, produces a peculiar pattern (Fig.\ref{f-SPEC}), whose shape slightly varies depending on the crystal dimensions, since these affect the $\gamma$ energy deposition containment.

Numerous studies of LYSO characteristics have been performed, as can be gathered in Ref.~\cite{r-MAO2}. 
The performance under $\gamma$-irradiation was thoroughly investigated~\cite{r-MAO},
$\gamma$-radiation effects have been shown to be small and dose rate dependent, with no recovery at ambient temperature. The performance for precision calorimetry~\cite{r-NOV} is such that attractive energy resolutions are achieved for photons between 200 and 500 MeV along with a time resolution as good as 150 ps. The capability to be grown in large ingots, needed for medium- and high-energy calorimetry, has been demonstrated~\cite{r-ZHUL} and the properties of such large samples have been studied~\cite{r-MAO2}.
The main concern about LYSO procurement remains the cost linked to raw materials, in that Lutetium is a rare rare-earth. Its performance when exposed to large fluxes of energetic hadrons, was so far unexplored. It is crucial for applications such as calorimetry at the HL-LHC and it is the subject of the study presented herein.
\section{Experimental setup}
\label{s-SETUP}
\begin{figure}[h]
\begin{center}
{\mbox{\includegraphics[width=12cm]{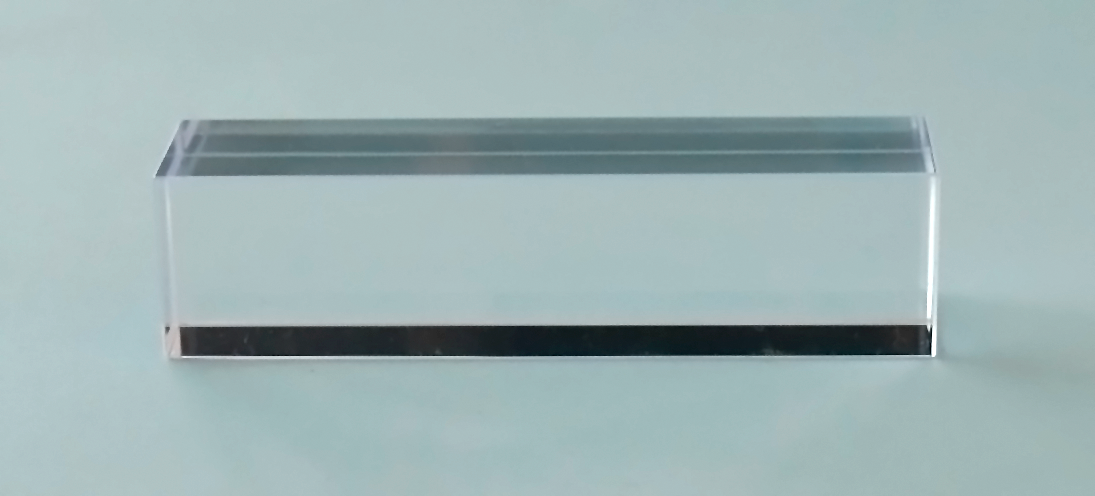}}}
\end{center}
\caption{One of the tested LYSO samples, showing excellent mechanical processing and optical surface treatment.}\label{f-CRY}
\end{figure}

The LYSO samples used for this test were produced by Saint-Gobain~\cite{r-GOB}, that commercializes LYSO under the name PreLude420, and by the Shanghai Institute of Ceramics (SIC)~\cite{r-SIC}. The samples have dimensions $25\times 25\times 100\;\mbox{mm}^3$, corresponding to 8.8 $X_0$ in length (Fig.~\ref{f-CRY}).

The irradiations of individual samples were performed at the IRRAD1 irradiation facility of the CERN PS T7 beam line~\cite{r-IRR}, with 24 GeV/c protons, up to different fluences representative for HL-LHC running conditions. The proton beam was widened in order to uniformly irradiate the whole crystal front face. The sample from SIC was exposed a first time in June 2009 to an integrated fluence
of 
 $\Phi^p_1= (8.85\pm 0.62) \times 10^{12}$\,cm$^{-2}$ with a flux $\phi^p_1 =(5.97\pm 0.42) \times 10^{12}$\,cm$^{-2}$\,h$^{-1}$.
 After measurements of damage, and recovery tracking over several months, the same sample was
 irradiated again in November 2010, up to an integrated fluence
of 
$\Phi^p_2= (7.24\pm 0.54) \times 10^{13}$\,cm$^{-2}$ with a flux  $\phi^p_2 =(3.82\pm 0.29) \times 10^{12}$\,cm$^{-2}$\,h$^{-1}$.
A proton irradiation of the LYSO sample from  St.~Gobain was performed in July 2011 to an integrated fluence
of 
$\Phi^p_3= (2.07\pm 0.16) \times 10^{13}$\,cm$^{-2}$ with a flux  $\phi^p_3 =(5.67\pm 0.43) \times 10^{12}$\,cm$^{-2}$\,h$^{-1}$.

The dosimetry for all irradiations was performed using the known activation cross-section of Aluminium foils placed in front of the sample during irradiation~\cite{r-LTNIM}. After irradiation, the samples were stored in the dark, at ambient temperature. The earliest measurements were taken when sample activation allowed for safe handling, according to the regulations for radioprotection.

\section{Visual observations}
\label{s-VIS}
While a slight loss of light transmission is barely visible through the naked eye in ambient light on the tested crystals, it is possible to
observe modifications due to the irradiation, by shining a LASER beam through the samples. 
In the photograph of Fig.~\ref{f-LAS}, one can observe how a beam of green laser light becomes visible when it traverses one of the proton-irradiated samples. This phenomenon was already observed in proton-irradiated 
Lead Tungstate crystals~\cite{r-LTNIM, r-FTNIM}, where it is very pronounced. As for Lead Tungstate, by placing a Polaroid filter between the right
half of the LYSO crystal and the camera, we observe cancellation, proving that the scattered light is polarized.
This feature has been linked, in our previous studies~\cite{r-LTNIM,r-FTNIM}, to Rayleigh scattering off small regions with different optical properties, as can be caused by disorder induced by hadron damage.
We conclude that at least a small fraction of the damage is due to light scattered off localized regions of damage, as it is the case for Lead Tungstate, where the presence of such regions, also called ``fission tracks'' in literature, have been recently visualized~\cite{r-FTNIM}.
This visualization in Lead Tungstate confirmed our understanding, first gained through indirect evidence~\cite{r-LTNIM}, of the mechanism at work:  the large energy deposits of fission fragments.

Lutetium, with Z=71, is right at the threshold for fission~\cite{r-THR}, and the relatively smaller probability for it to happen might explain why the effect is less pronounced here than what observed in PbWO$_4$~\cite{r-FTNIM}.
\begin{figure}[h]
\begin{center}
{\mbox{\includegraphics[width=12cm]{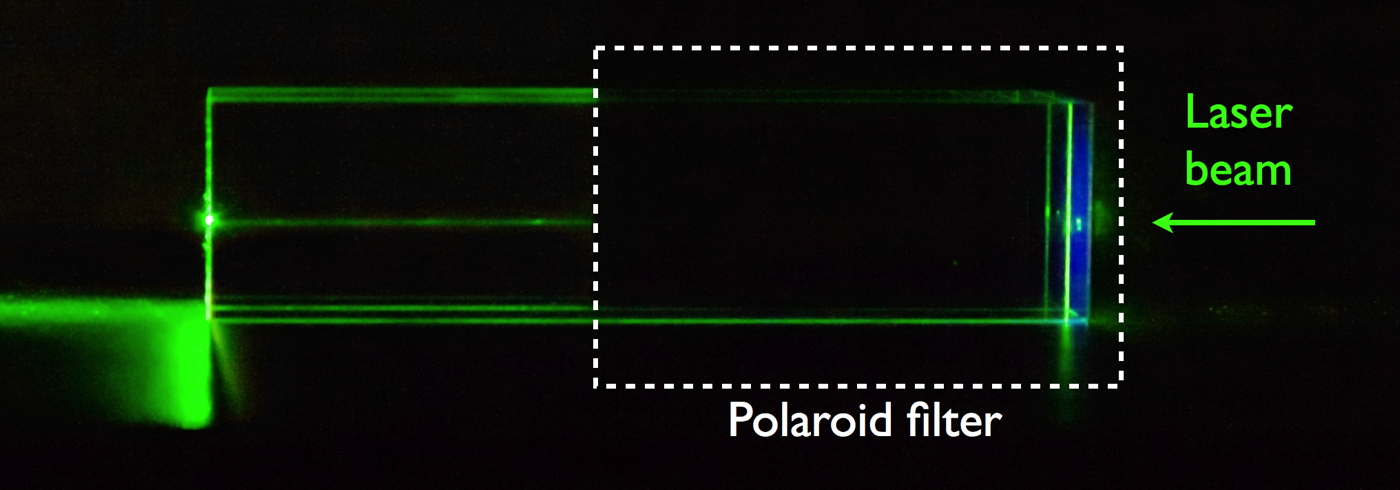}}}
\end{center}
\caption{A faint beam reveals the green Laser light shone through a proton-irradiated LYSO crystal and thus indicates the presence of scatterers. Cancellation by a Polaroid filter indicates that the scattered light is polarized.}\label{f-LAS}
\label{f-LAS}
\end{figure}

\section{Changes in Light Transmission}
\label{s-LT}
\begin{figure}[h]
\begin{center}
 \begin{tabular}[h]{cc}
{\mbox{\includegraphics[width=80mm]{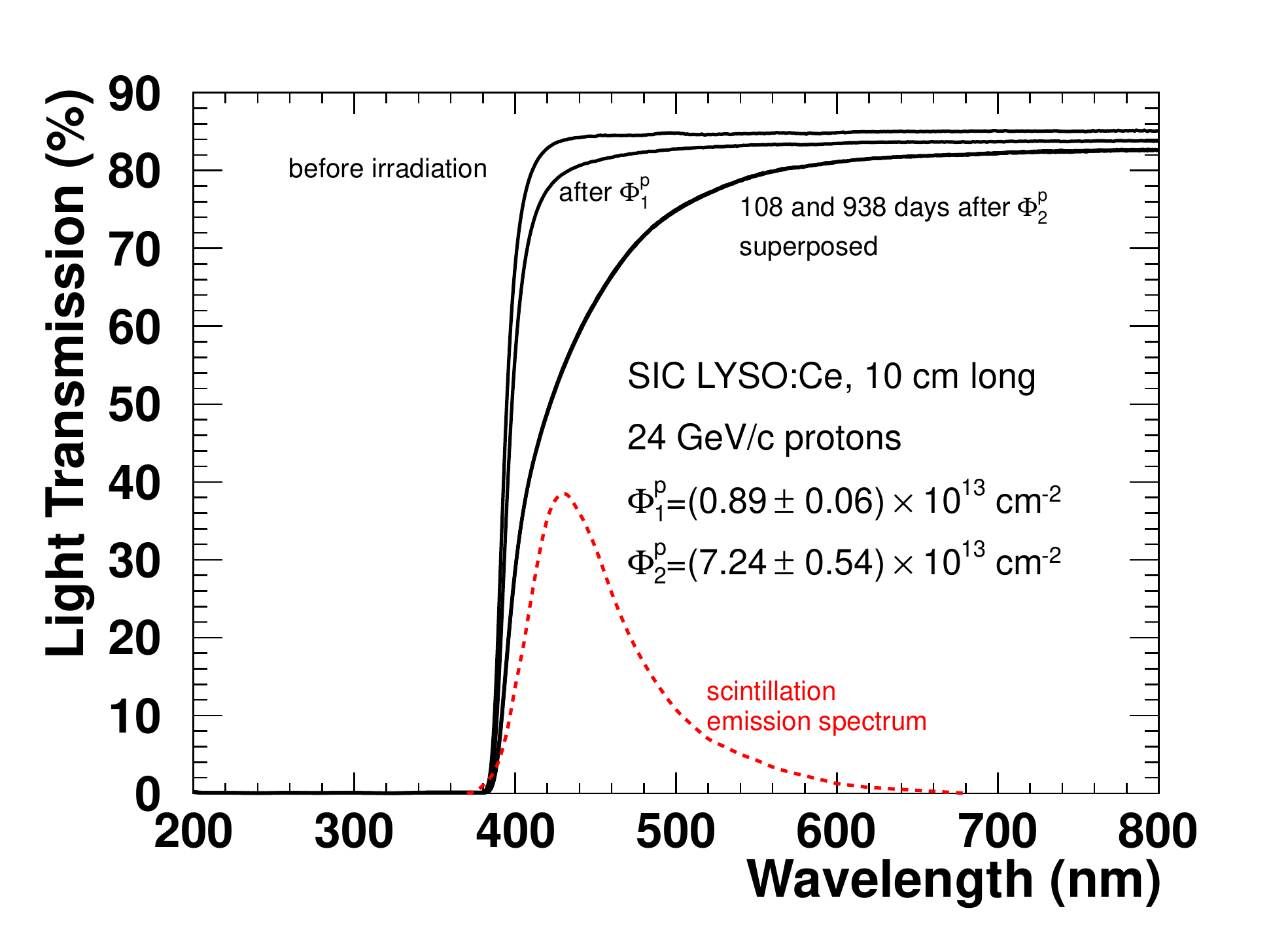}}} &
{\mbox{\includegraphics[width=80mm]{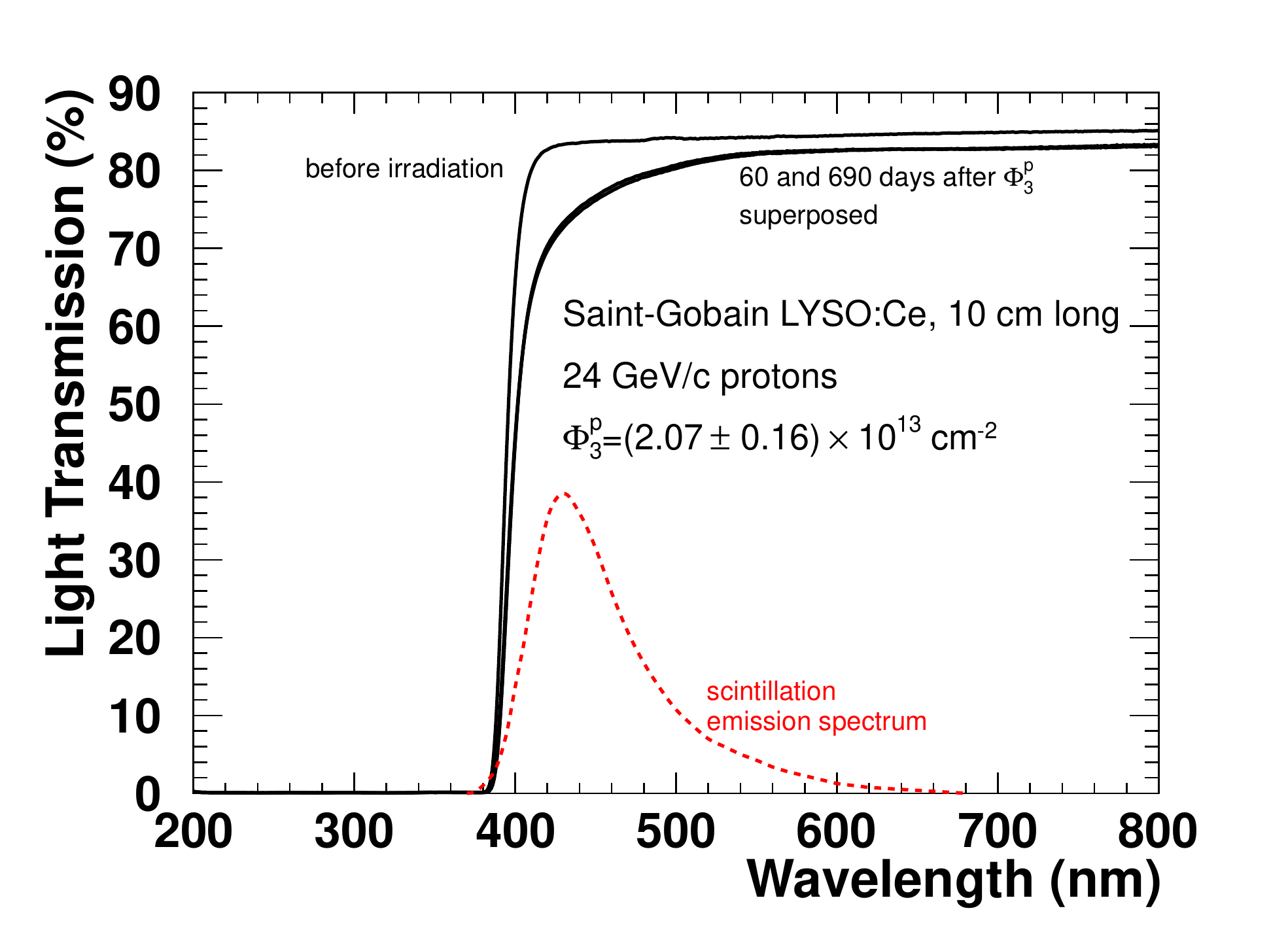}}}
   \end{tabular}
\end{center}
\caption{Light transmission as a function of wavelength, measured longitudinally through the 100 mm length of the crystals. Left: values before irradiation and at two different times after each irradiation of the SIC sample along with the scintillation emission spectrum from~\cite{r-MAO2}\label{f-LT}. Right: values before and at two different times after irradiation of the St.~Gobain sample. }
\end{figure}

To quantify the loss of light transmission induced by hadrons, we have measured it as a function of wavelength, in 1 nm steps, 
with a high-precision Lambda900 spectrophotometer from Perkin-Elmer~\cite{r-PE}. The resulting transmission curves before irradiation and at selected times after irradiation, are visible in Fig.~\ref{f-LT}, uncorrected for Fresnel losses.
One can observe how the loss of transmission is smooth, and how the typical dips related to color centers are absent. This indicates that color center formation --- as typical for ionizing damage --- is not the main damage mechanism here. Transmission curves taken at different intervals after irradiation are superposed, indicating that the loss does not recover at room temperature, over a time span of one year or longer.
For this reason, in later plots we do not quote any further the time after irradiation of the measurements.
\begin{figure}[b]
\begin{center}
{\mbox{\includegraphics[width=10cm]{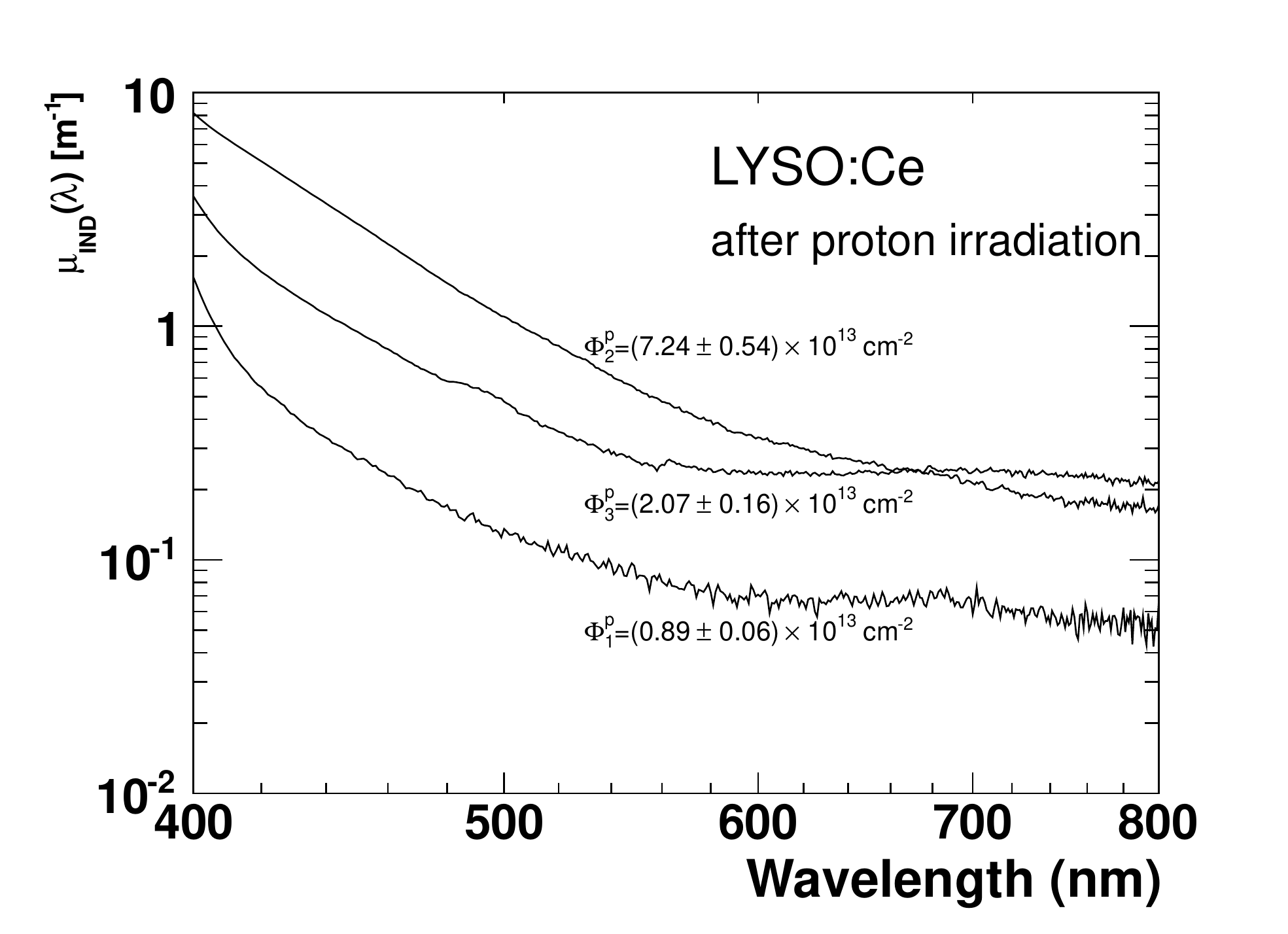}}}
\end{center}
\caption{Induced absorption as a function of wavelength, measured longitudinally through the 100 mm length of the SIC crystal after each of the two irradiations with fluences $\Phi^p_1$ and  $\Phi^p_2$, and for the St.~Gobain crystal after irradiation with fluence $\Phi^p_3$. \label{f-MULAMBDA}}
\end{figure}

We have determined the induced absorption $\mu_{IND}(\lambda)$  as a function of wavelength $\lambda $, that is defined as
\begin{equation}
\frac{LT(\lambda )}{LT\it{0}(\lambda )} = \exp^{-\mu_{IND}(\lambda )L}
\end{equation}
with {\em L} the crystal length and {\em LT0 (LT)} the light transmission before (after) irradiation. The comparison in Fig.~\ref{f-MULAMBDA}, between the data for the three fluences, indicates proportionality between damage and $\Phi^p$. A small, unexplained deviation from a scaling behavior is observed for the St.~Gobain sample for $\lambda > 550$ nm, that is however far from the range of wavelengths relevant for the scintillation light collection.
The smoothness of the data curves confirms the absence of dominant color centers. Their slope is shallower than for Lead Tungstate~\cite{r-LTNIM}, where a $\lambda^{-4}$ behavior was observed,
typical for Rayleigh scattering. This latter observation indicates that, although the creation of some scattering centers through hadrons must have occurred, as illustrated in Section~\ref{s-VIS}, their presence is by far not the dominant mechanism to explain the loss of light transmission.

\begin{figure}
\begin{center}
{\mbox{\includegraphics[width=10cm]{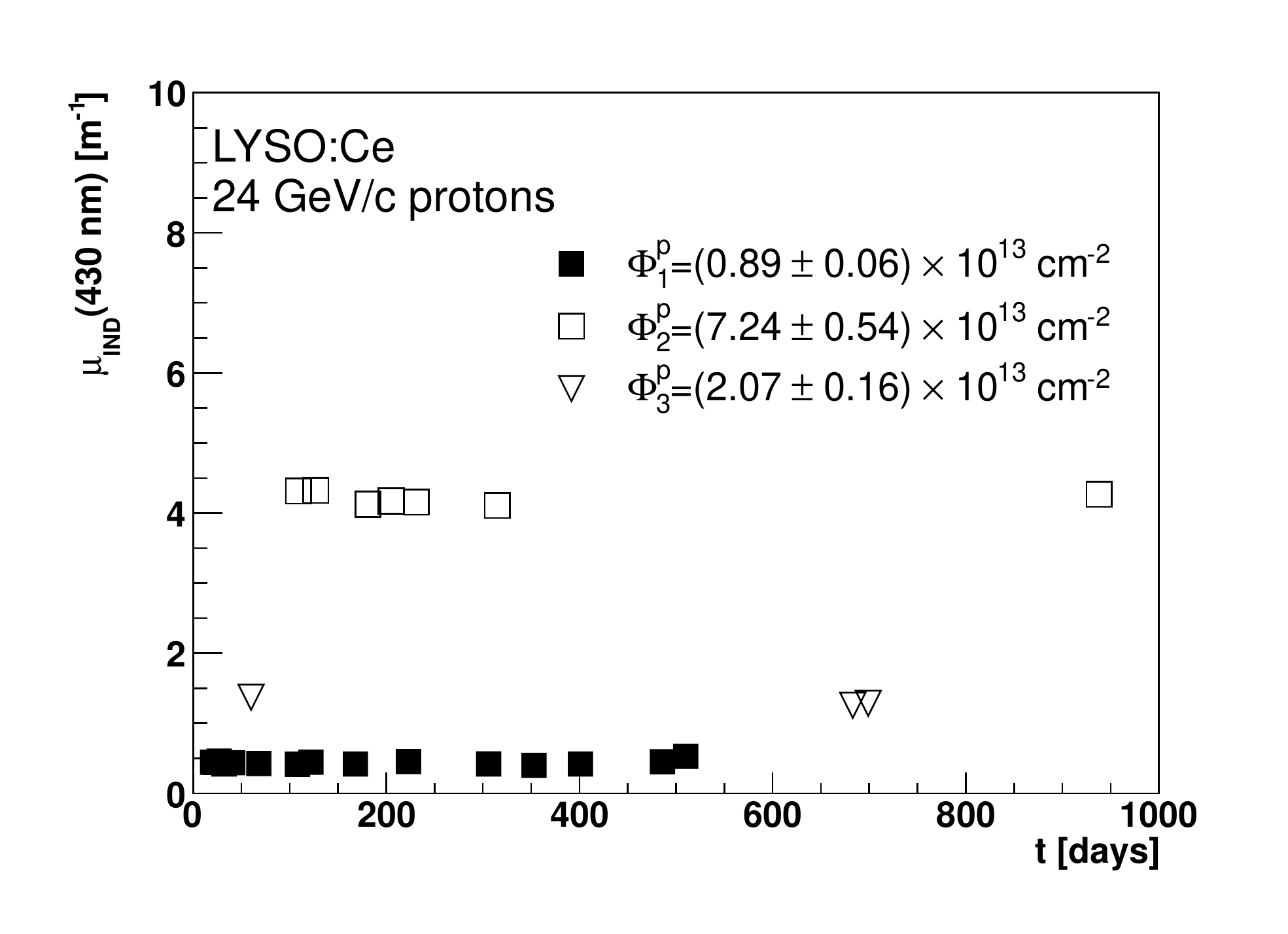}}}
\end{center}
\caption{Induced absorption at the peak-of-emission wavelength measured at different time intervals after each of the three irradiations. \label{f-MUTIME}}
\end{figure}

Measurements of the induced absorption at different intervals after irradiation are presented in Fig.~\ref{f-MUTIME},
where values are shown at 430 nm, the peak-of-emission wavelength, for different intervals after irradiation.
No significant recovery is observed, thus implying a long-term damage that is permanent at room temperature,
when observed over several months and even years after irradiation.

\begin{figure}
\begin{center}
{\mbox{\includegraphics[width=12cm]{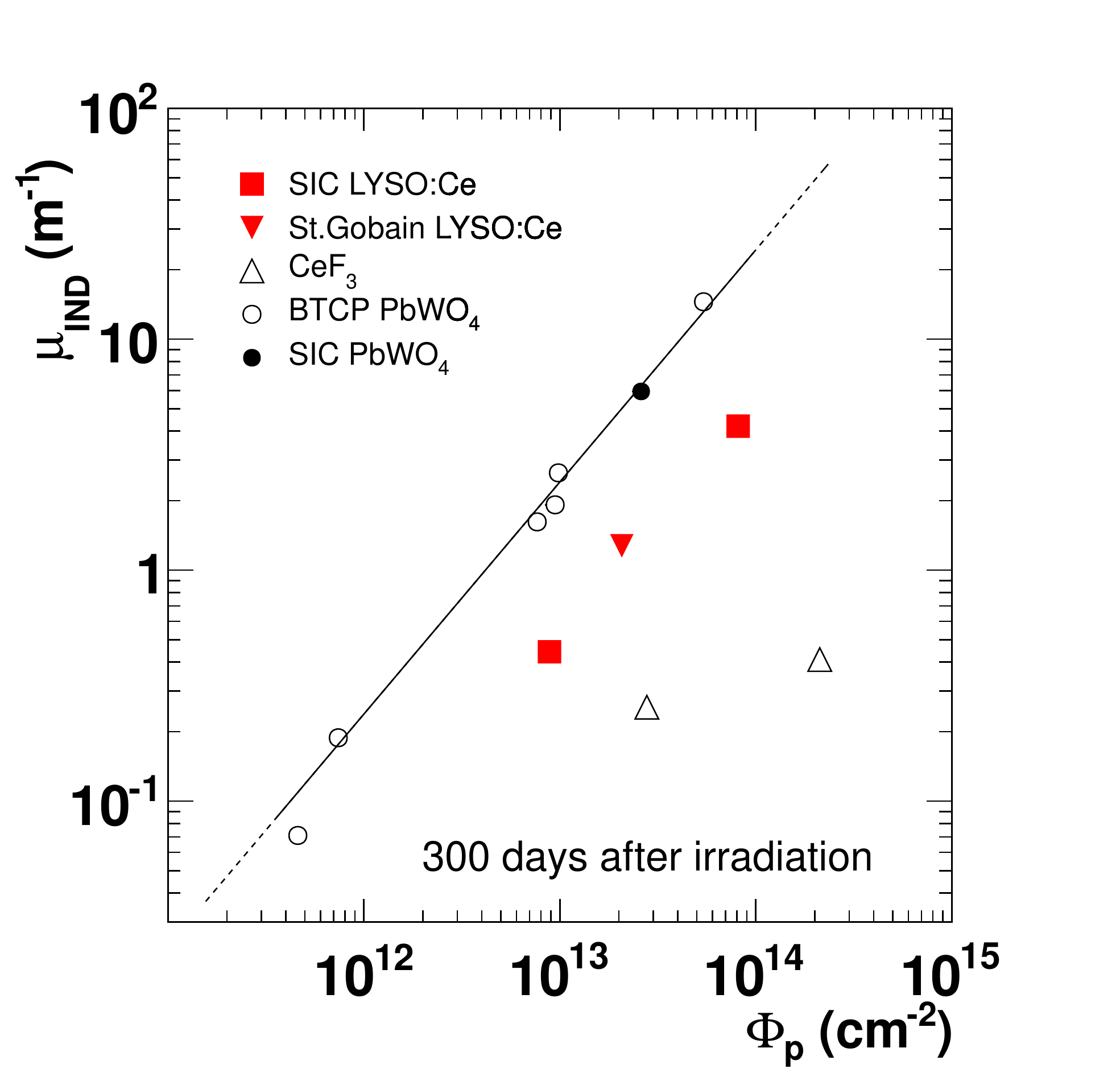}}}
\end{center}
\caption{Induced absorption at the peak-of-emission wavelength for LYSO:Ce crystals as a function of
integrated proton fluence, in comparison to Lead Tungstate~\cite{r-LTNIM} and Cerium Fluoride~\cite{r-CEF3}\label{f-MUCUMU}.}
\end{figure}

For the LYSO:Ce evaluation as a possible alternative scintillator to be used at the HL-LHC, it is useful to compare its
level of damage as a function of integrated proton fluence with the ones of Lead Tungstate and Cerium Fluoride. This is done in
Fig.~\ref{f-MUCUMU}, where the central data point for LYSO:Ce corresponds to the crystal produced by St.~Gobain, the other ones to the two irradiation tests of the SIC crystal. 
The plot shows how the damage in LYSO:Ce is cumulative, similarly to what observed for Lead Tungstate. However, the amplitude of induced absorption at the peak-of-emission wavelength is a factor 5 smaller than 
the value measured in Lead Tungstate.  The values for Cerium Fluoride instead recover and their value depends on the time elapsed after irradiation~\cite{r-CEF3}. The observed trend indicates that hadron effects in LYSO do not depend on doping level, growth parameters and fine details of its production, as would be the case for different manufacturers.

\section{Radioactivation}
\label{s-RA}

The remnant radioactivity after hadron irradiation is relevant in case a human intervention is needed after the detector containing such crystals has been operated. Since the fluences in this test have been delivered to the samples over few hours, the absolute values measured here are not relevant for what would be observed in situ. However, the comparison between different crystal types is crucial to anticipate the expected exposure.

As a first step, the induced ambient-dose equivalent rate  (``dose'') ${\dot{H}^*(10)_{\rm ind}}$  has been measured  with a dose-rate meter AUTOMESS 6150AD5~\cite{r-AUTOMESS} at 5.7 cm distance from the middle of the SIC sample at various times after irradiation. The measurements are compared  with results from a
FLUKA simulation (version 2011.2b.3, \cite{r-FLUKA},\cite{r-FLUKA2}) , after rescaling to $\Phi_p = 10^{13}\; \mathrm{p/cm^2}$. 
The simulation setup was validated with former FLUKA studies for Lead Tungstate crystals \cite{r-LTNIM} and adapted to the current experimental setup.
The proton-induced hadronic shower within the crystal can lead to very forward-directed particles hitting the beam dump of the irradiation zone in the T7 facility. 
To include possibly backscattered neutrons, the back wall of the irradiation zone was included in the geometry setup, whereas the side walls were neglected.
In the simulation, the LYSO crystal was exposed to a square-shaped, uniform beam ($5\times 5\,\mathrm{cm}^2$) of 24 GeV/c  protons with an integrated fluence of $\Phi_p = 10^{13}\; \mathrm{p/cm^2}$.
To enable the physics models optimized for activation studies in FLUKA, precise thresholds for the particle transport were applied, which include a low energy neutron transport down to thermal energies and the coalescence and the evaporation of heavy fragments. 
The ambient dose equivalent rate ${\dot{H}^*(10)_{\rm ind}}$ of the single crystal was simulated in an area with low background and recorded at a distance of 5.7 cm from the middle of the crystal, according to the experimental setup. For the calculation, the respective conversion coefficients were used \cite{convfactors}. 

\begin{figure}[b]
\begin{center}
{\mbox{\includegraphics[width=12cm]{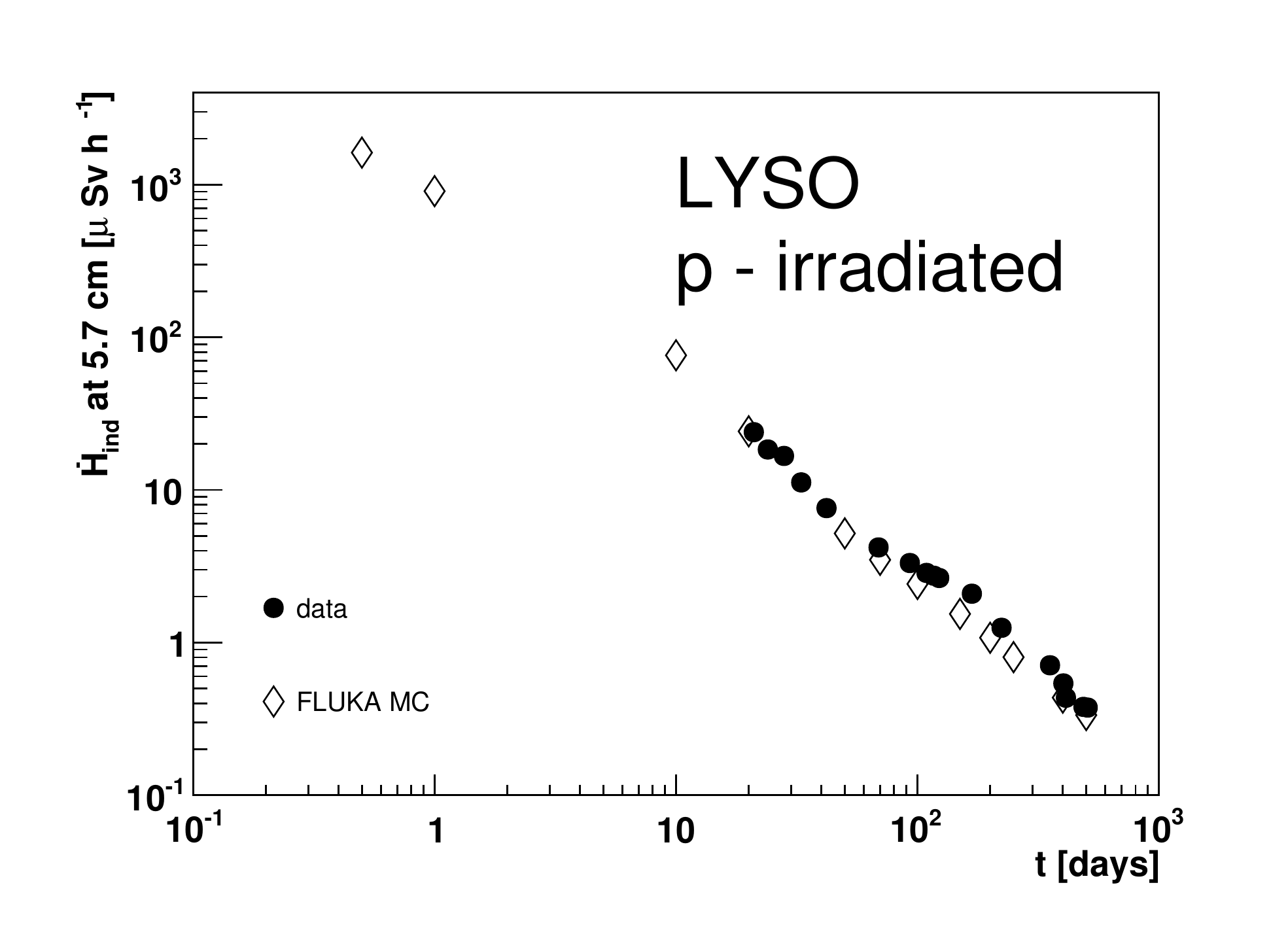}}}
\end{center}
\caption{Induced ambient-dose equivalent rate as a function of time after the first irradiation, measured 5.7 cm from the side of the crystal, compared to FLUKA simulations\label{f-EQDOSE}.}
\end{figure}

The comparison of the measured residual dose rate with the FLUKA simulation results can be seen in Fig.~\ref{f-EQDOSE}.
A very good agreement is observed, between FLUKA simulations and measurements, over one year and over two orders of magnitude in dose, with the simulation results falling slightly on the low side. 
Uncertainties in the comparison can arise from the limited knowledge of Lutetium cross-sections in FLUKA, and from non-uniformities of the beam intensity during irradiation combined with the self-shielding of the crystal.
The comparison benchmarks the reliability and predictive power of FLUKA for such LYSO crystals.\\

\begin{figure}
\begin{center}
{\mbox{\includegraphics[width=10cm,trim=0mm 10mm 0mm 0mm , clip]{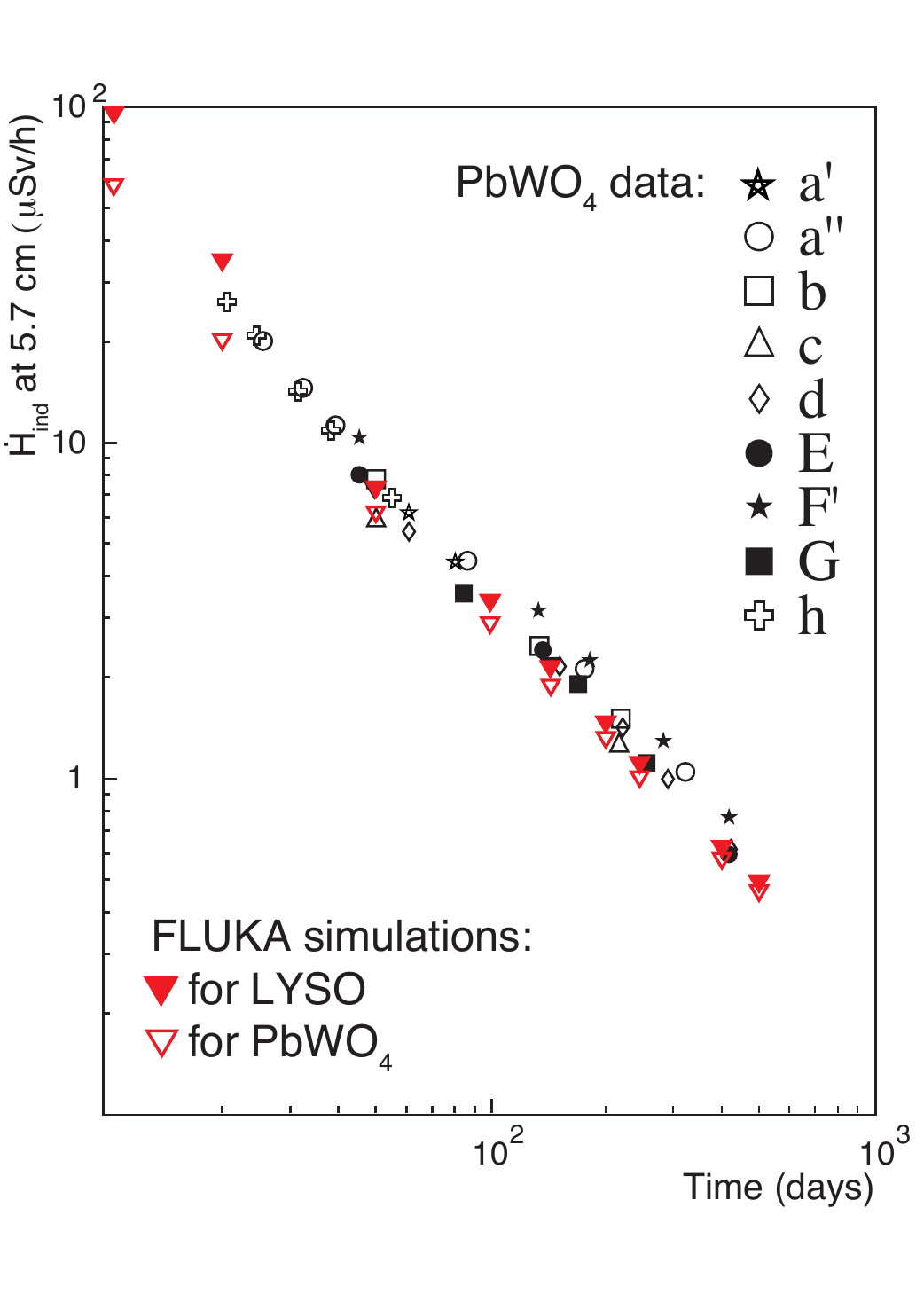}}}
\end{center}
\caption{Induced ambient-dose equivalent rate as a function of time after irradiation from FLUKA simulations for LYSO compared to simulations and to measurements for Lead Tungstate from~\cite{r-LTNIM} (all crystals 26 $X_0$ long).\label{f-FULLSIZE}}
\end{figure}

In a second step, the FLUKA simulations have been extended to full-size crystals, as could be used for a homogenous calorimeter at the HL-LHC collider. In this case, it is relevant to compare the expected remnant radioactivity among crystals of similar dimensions in terms of radiation lengths, typical for such a calorimeter.
The crystal length was 26~$X_0$, that corresponds to 30 cm for LYSO and 23 cm for PbWO$_4$, while we kept the the same granularity ($24\times 24\,\mathrm{mm}^2$). The simulation results for Lead Tungstate are compared to existing measurements \cite{r-LTNIM}. Accordingly, the crystals were assumed to be proton-irradiated with a 20 GeV/c square-shaped beam profile for an integrated fluence of $\Phi_p = 10^{13}\;\mathrm{p/cm^2}$.  The ambient dose equivalent rate ${\dot{H}^*(10)_{\rm ind}}$ for each crystal was recorded as described  above and is depicted in Fig.~\ref{f-FULLSIZE}.
The FLUKA simulations for LYSO show a remnant dose similar to the one of Lead Tungstate, with the simulation results for Lead Tungstate being roughly a factor 2 below the measurements. This is an indication that LYSO might become sightly more radioactive than Lead Tungstate in a $26\;\mbox{X}_0$ deep calorimeter.

\section*{Conclusions}
Exposure to energetic hadrons of LYSO crystal samples causes a loss in light transmission, that is cumulative, with no sign of recovery over time at room temperature. After an irradiation with  24 GeV/c protons up to a fluence $\sim7\times10^{13}/\mathrm{cm}^2$, an induced absorption coefficient of $4\;\mbox{m}^{-1}$ is observed at the peak-of-scintillation-emission wavelength of 430 nm, five times smaller than what observed for Lead Tungstate exposed to the same irradiation conditions. Proton irradiations at different fluences, and for different sample manufacturers, give the same damage ratio compared to Lead Tungstate. This indicates that hadron-induced damage is independent from the manufacturer, and thus also from fine growth and composition details. The expected remnant radioactivity is of the same order of magnitude as for Lead Tungstate.

\clearpage
\end{document}